\documentclass[12pt]{article}
\begin{document}
\title{Quantum Mechanics and Multiply Connected Spaces}
\author{B.G. Sidharth\\
International Institute for Applicable Mathematics \& Information Sciences\\
Hyderabad (India) \& Udine (Italy)\\
B.M. Birla Science Centre, Adarsh Nagar, Hyderabad - 500 063 (India)}
\date{}
\maketitle
\begin{abstract}
It is well known that the difference between Quantum Mechanics and Classical Theory appears most crucially in the non Classical spin half of the former theory and the Wilson-Sommerfelt quantization rule. We argue that this is symptomatic of the fact that Quantum Theory is actually a theory in multiply connected space while Classical Theory operates in simply connected space.
\end{abstract}
\section{Introduction}
John Wheeler had stressed that the divide between Classical and Quantum Theory lies in the spin half (of Fermions) \cite{mwt}. This half integral spin gives rise to such non Classical and purely Quantum Mechanical results as the anomalous gyromagnetic ratio of the electron $(g=2)$. We will now argue that the non Classical half integral spin feature arises from the multiply connected nature of Quantum Spacetime, and it is this which distinguishes Quantum Mechanics from Classical Theory. Specifically, we will argue that the usual space $R$ with a compactified space $S^1$, in $R \times S^1$ reproces Quantum Mechanical spin. On the other hand, spacetime is simply connected in Classical theory.
\section{Multiply Connected Space on Spin}
Let us start by reviewing Dirac's original derivation of the Monopole (Cf.ref.\cite{dirac2}). He started with the wave function
\begin{equation}
\psi = Ae^{\imath \gamma},\label{e1}
\end{equation}
He then considered the case where the phase $\gamma$ in (\ref{e1}) is non integrable. In this case (\ref{e1}) can be rewritten as
\begin{equation}
\psi = \psi_1 e^{\imath S},\label{e2}
\end{equation}
where $\psi_1$ is an ordinary wave function with integrable phase, and further, while the phase $S$ does not have a definite value at each point, its four gradient viz., 
\begin{equation}
K^\mu = \partial^\mu S\label{e3}
\end{equation}
is well defined. We use temporarily natural units, $\hbar = c = 1$. Dirac then goes on to identify $K$ in (\ref{e3}) (except for the numerical factor $hc/e$) with the electromagnetic field potential, as in the Weyl gauge invariant theory.\\
Next Dirac considered the case of a nodal singularity, which is closely related to what was later called a quantized vortex (Cf. for example ref.\cite{vasu}). In this case a circuit integral of a vector as in (\ref{e3}) gives, in addition to the electromagnetic term, a term like $2 \pi n$, so that we have for a change in phase for a small closed curve around this nodal singularity,
\begin{equation}
2 \pi n + e \int \vec B \cdot d \vec S\label{e4}
\end{equation}
In (\ref{e4}) $\vec B$ is the magnetic flux across a surface element $d \vec S$ and $n$ is the number of nodes within the circuit. The expression (\ref{e4}) directly lead to the Monopole in Dirac's formulation.\\
Let us now reconsider the above arguments in terms of recent developments. The Dirac equation for a spin half particle throws up a complex or non Hermitian position coordinate \cite{bgsfpl,fpl2}. Dirac identified the imaginary part with zitterbewegung effects and argued that this would be eliminated once it is realized that in Quantum Mechanics, spacetime points are not meaningful and that on the contrary averages over intervals of the order of the Compton scale have to be taken to recover meaningful physics \cite{dirac}. Over the decades the significance of such cut off space time intervals has been stressed by T.D. Lee and several other scholars as noted earlier \cite{cu,x2,x5,x9}. Indeed with a minimum cut off length $l$, it was shown by Snyder that there would be a non commutative but Lorentz invariant spacetime structure. At the Compton scale we would have \cite{uof},
\begin{equation}
[x,y] = 0(l^2)\label{e5}
\end{equation}
and similar relations.\\
In fact starting from the Dirac equation itself, we can deduce directly the non commutativity (\ref{e5}) (Cf.refs.\cite{bgsfpl,fpl2}).\\
Let us now return to Dirac's formulation of the monopole in the light of the above comments. As noted above, the non integrability of the phase $S$ in (\ref{e2}) gives rise to the electromagnetic field, while the nodal singularity gives rise to a term which is an integral multiple of $2 \pi$. As is well known \cite{rr12} we have
\begin{equation}
\vec \nabla S = \vec p\label{e6}
\end{equation}
where $\vec p$ is the momentum vector. When there is a nodal singularity, as noted above, the integral over a closed circuit of $\vec p$ does not vanish. In fact in this case we have a circulation given by
\begin{equation}
\Gamma = \oint \vec \nabla S \cdot d \vec r = \hbar \oint dS = 2 \pi n\label{e7}
\end{equation}
It is because of the nodal singularity that though the $\vec p$ field is irrotational, there is a vortex - the singularity at the central point associated with the vortex makes the region multiply connected, or alternatively, in this region we cannot shrink a closed smooth curve about the point to that point. In fact if we use the fact as seen above that the Compton wavelength is a minimum cut off, then we get from (\ref{e7}) using (\ref{e6}), and on taking $n = 1$,
\begin{equation}
\oint \vec \nabla S \cdot d\vec r = \int \vec p \cdot d \vec r = 2\pi mc \frac{1}{2mc} = \frac{h}{2}\label{e8}
\end{equation}
$(l = \frac{\hbar}{2mc}$ is the radius of the circuit and $\hbar = 1$ in the above natural units). In other words the nodal singularity or quantized vortex gives us the mysterious Quantum Mechanical spin half (and other higher spins for other values of $n$). In the case of the Quantum Mechanical spin, there are $2 \times n/2 + 1 = n + 1$ multiply connected regions, exactly as in the case of nodal singularities. Indeed in the case of the Dirac wave function, which is a bi-spinor $\left(\begin{array}{ll}
\Theta \\ \phi
\end{array}
\right),$ 
as is known \cite{bd}, far outside the Compton wavelength, it is the usual spinor $\Theta$, preserving parity under reflections that predominates, whereas at and near the Compton scale it is the spinor $\phi$ which predominates, where under a reflection $\phi$ goes over to $- \phi$.\\
The multiply connected nature of Quantum Spacetime and spin half can be seen to emerge from (\ref{e5}). In fact given (\ref{e5}), it is immediately seen that we cannot shrink a close circuit about the point $x,y$ to a single point.\\
One can argue that starting from (\ref{e5}) it is possible to obtain directly Quantum Mechanical spin and the Dirac representation. It has been shown in detail \cite{bgscsf,bgsnc} that under a time elapse transformation of the wave
function, (or, alternatively, as a small scale transformation),
\begin{equation}
| \psi' > = U(R)| \psi >\label{e9}
\end{equation}
we get
\begin{equation}
\psi' (x_j) = [1 + \imath \epsilon (\imath x_j \frac{\partial}{\partial x_j}) + 0 (\epsilon^2)] \psi
(x_j)\label{e10}
\end{equation}
Equation (\ref{e10}) has been shown to lead to the Dirac equation
when $\epsilon$ is the Compton time. A quick way to see this is as follows: At the Compton scale we have,
$$|\vec {L} | = | \vec {r} \times \vec {p} | = | \frac{\hbar}{2mc} \cdot mc| = \frac{\hbar}{2},$$
that is, we get the Quantum Mechanical spin. Next, we can easily verify, that the choice,
$$t = \left(\begin{array}{ll}
1 \quad 0\\
0 \quad -1
\end{array}
\right), \vec {x} = \left(\begin{array}{ll}
0 \quad \vec {\sigma}\\
\vec {\sigma} \quad 0
\end{array}
\right)$$
provides a representation for the coordinates in (3), apart from scalar factors. As can be seen, this is also a representation of the Dirac matrices. Substitution of the above in (\ref{e10}) leads to the Dirac equation
$$(\gamma^\mu p_\mu - mc^2) \psi = 0$$
because
$$E\psi = \frac{1}{\epsilon}\{\psi' (x_j) - \psi (x_j)\}, \quad E = mc^2,$$
where $\epsilon = \tau$ (Cf.ref.\cite{wolf}).\\
Indeed, as noted, Dirac himself had
realized that his electron equation needed an average over spacetime
intervals of the order of the Compton scale to remove
zitterbewegung effects and give meaningful physics. This again is symptomatic of an underlying fuzzy
spacetime described by a noncommutative space time geometry
(\ref{e7}) or (4) \cite{sakharov}.\\
The point here is that under equation (xxx), the coordinates
$x^\mu \to \gamma^{(\mu)} x^{(\mu)}$ where the brackets with the
superscript denote the fact that there is no summation over the
indices.  Infact, in the theory of the Dirac equation it is well
known \cite{bade}that,
\begin{equation}
\gamma^k \gamma^l + \gamma^l \gamma^k = - 2g^{kl}I\label{e11}
\end{equation}
where $\gamma$'s satisfy the usual Clifford algebra of the Dirac
matrices, and can be represented by
\begin{equation}
\gamma^k = \sqrt{2} \left(\begin{array}{ll}
0 \quad \sigma^k \\
\sigma^{k*} \quad 0
\end{array}\right)\label{e12}
\end{equation}
where $\sigma$'s are the Pauli matrices. As noted years ago by Bade and
Jehle (Cf.ref.\cite{bade}), we could take the $\sigma$'s or
$\gamma$'s in (\ref{e11}) and (\ref{e12}) as the components of a
contravariant world vector, or equivalently we could take them to
be fixed matrices, and to maintain covariance, to attribute new
transformation properties to the wave function, which now becomes
a spinor (or bi-spinor). This latter has been the traditional
route, because of which the Dirac wave function has its
bi-spinorial character. In this latter case, the coordinates
retain their usual commutative or point character. It is only when we
consider the equivalent former alternative, that we return to the
noncommutative
geometry (\ref{e5}).\\
That is, in the usual commutative spacetime the Dirac spinorial
wave functions conceal the noncommutative
character (\ref{e5}).
\section{Discussion}
1. We consider a Hydrogen like atom in two dimensional space, for which the Schrodinger equation is given by \cite{ho,schiff}
\begin{equation}
-\frac{\hbar^2}{2\mu} \left[\frac{1}{r} \frac{\partial}{\partial r} \left(r\frac{\partial}{\partial r}\right) + \frac{1}{r^2}\frac{\partial^2}{\partial \phi^2}\right] \psi(r,\phi) - \frac{Ze^2}{r}\phi(r,\phi) = E\psi(r,\phi)\label{e13}
\end{equation}
As is well known the energy spectrum for (\ref{e13}) is given by
\begin{equation}
E = -\frac{Z^2e^4\mu}{2\hbar^2(n+m+\frac{1}{2})^2}\label{e14}
\end{equation}
If we require that (\ref{e14}) be identical to the Bohr spectrum, then $m$ should be a half integer, which also means that the configuration space is multiply connected. In the simplest case of a doubly connected space, we are dealing wityh $R^2 \times S^1$, where $S^1$ is a compactified space, generally considered to be a Kaluza-Klein space. However we would like to point out the following: The energy is given by
\begin{equation}
E = \left(k^2 + \frac{S^2}{\rho^2}\right)\frac{\hbar^2}{2\mu}\label{e15}
\end{equation}
In (\ref{e15}) there is an additional grounds state energy $E = S^2 \hbar^2 /2\mu \rho^2$, where $\mu$ is the reduced mass and $\rho$ is the radius of the compactified circle $S^1$. If $\rho$ were to be the Planck length as in the Kaluza-Klein theory, then this extra energy becomes very large and is generally taken to be unobservable. On the other hand if $\rho$ is taken to be the Compton wavelength as in our earlier discussion, then the above extra ground energy, as can be easily verified is of the same orderas the usual energy. In any case it can be seen that the Quantum Mechanical spin is a symptom of the multiply connected nature of Quantum spacetime, even in this non relativistic example. We remark that, as is well known, in (\ref{e14}), we can continue to take integral values of the momentum $m$, provided, to the Coulomb potential energy an additional energy
\begin{equation}
\Delta E = \hbar \left(\sqrt{E/2\mu}\right)/r\label{e16}
\end{equation}
is added. It is immediately seen that if in (\ref{e16}) $r$ is of the order of the Compton wavelength, which is also $\sim e^2 / mc^2,$ then we recover $e^2$.To put it another way, if there was no Coulomb interaction in the conventional theory, then this additional contribution shows up as the Coulomb field.  This suggests the origin of the fundamental charge itself from topological conditions.\\
2. In Dirac's theory of displacement operators \cite{dirac} the operator $d_x \equiv \frac{d}{dx}$ is
a purely imaginary operator, and is given by
$$\delta x (d_x + \bar d_x) = \delta x^2 d_x \bar d_x = 0$$
if
$$0 (\delta x^2) = 0$$
as is tacitly assumed. However if
\begin{equation}
0 (\delta x^2) \ne 0\label{e17}
\end{equation}
then the operator $d_x$ becomes complex, and therefore, also the
momentum operator, $p_x \equiv \imath \hbar d_x$ and the position
operator. In other words if (\ref{e17}) holds good then we have to
deal with complex or non-Hermitian coordinates. The implication of
this is that (Cf.\cite{annales} for details) spacetime becomes non-
commutative as seen above.\\
In any case here is the mysterious origin of the complex
coordinates and spin. It is the complex coordinates that lead from the Coulomb potential to the electromagnetic part of the
Kerr-Newman metric and the electron's field including the
anomalous gyro magnetic ratio which are symptomatic of the
electron's spin \cite{newman}. It also means that the naked singularity is
shielded by the fuzzy spacetime (Dirac's original averages over
the zitterbewegung interval) or equivalently the noncommutative
geometry (\ref{e5}) (Cf. also \cite{madore}). Indeed, if we remember that $\vec{\nabla} S$ in (\ref{e8}) gives the momentum $\vec{p}$, we can see that (\ref{e8}) is an expression of the Wilson-Sommerfeld quantization rule \cite{afdb}. What all this means is that the presence of a Fermion in usual simply connected space tantamounts to making the space multiply connected - like a hole in a sheet.


\begin{thebibliography}{99}
\bibitem {mwt} Misner, C.W., Thorne, K.S., and Wheeler, J.A.,  "Gravitation", W.H. Freeman,
San Francisco, 1973, p.819ff.
\bibitem {dirac2} Dirac, P.A.M.,  Proc. Roy. Soc. \underline{A 133}, 1931, pp.60ff.
\bibitem {vasu} Vasudevan, R.,  ``Hydrodynamical Formulation of Quantum Mechanics'', in ``Perspectives in Theoretical Nuclear Physics'', Ed. Srinivas Rao, K., and Satpathy L., Wiley Eastern, New Delhi, 1994, pp.216ff.
\bibitem {bgsfpl} Sidharth, B.G., Foundation of Physics Letters, 15 (5), 2002, 501ff.
\bibitem {fpl2} Sidharth, B.G., Foundation of Physics Letters, 16 (1), 2003, pp.91-97.
\bibitem {dirac} Dirac, P.A.M., in ``The Principles of Quantum Mechanics'', Clarendon Press, Oxford, 263, 1958.
\bibitem {cu} Sidharth, B.G., ``The Chaotic Universe: From the Planck to the Hubble Scale'', Nova Science Publishers, Inc., New York, 2001.
\bibitem {x2} Kardyshevskii, V.G.,  Translated from Doklady Akademii Nauk SSSR, Vol.147, No.6, December 1962, p.1336-1339.
\bibitem {x5} Bombelli, L.,  Lee, J., Meyer, D., and  Sorkin, R.D.,  Physical Review Letters, Vol.59, No.5, August 1987, p.521-524.
\bibitem {x9} Lee, T.D.,  Physics Letters, Vol.122B, No.3,4, 10 March 1983, p.217-220.
\bibitem {uof} Sidharth, B.G., ``The Universe of Fluctuations'', Springer, Berlin, 2005.
\bibitem {rr12} Sidharth, B.G., Chaos, Solitons and Fractals, 12(1), 2001,
173-178.
\bibitem {bd} Bjorken, J.D.,  and  Drell, S.D., ``Relativistic
Quantum Mechanics'', Mc-Graw Hill, New York, 1964, 24.
\bibitem {bgscsf} Sidharth, B.G., Chaos, Solitons and Fractals, 11, (8), 2000, 1269-1278.
\bibitem {bgsnc} Sidharth, B.G. Nuovo Cimento, 117B, (6), 2002, pp.703ff.
\bibitem {wolf}  Wolf, C.,  Hadronic Journal, Vol.13, 1990, p.208-210.
\bibitem {sakharov} Sidharth, B.G., Found.Phys.Lett., 16 (1), 2003, pp.91-97.
\bibitem {bade} Bade, W.L.,  and Herbert Jehle,  Reviews of Modern Physics, {\bf 25} (3), 1952.
\bibitem {ho} Ho, Vu B., Morgan J Michael, J.Phys.A: Math. Gen. {\bf 29}, (1996), 1497-1510.
\bibitem {schiff} Schiff, L.I.,  ``Quantum Mechanics'', McGraw Hill, London, 1968.
\bibitem {annales} Sidharth, B.G.,  Annales de la Fondation Lois de Broglie, Volume 27 No.2, 2002, 333-342.
\bibitem {newman} Newman, E.T.,  J.Math.Phys., 14 (1), 1973, p.102.
\bibitem {madore} Madore, J., ``An Introduction to Non-Commutative Differential Geometry'', Cambridge University Press, Cambridge, 1995.
\bibitem {afdb} Sidharth, B.G., Annales Fondation L De Broglie, 29 (3), 2004, 1.
\end{thebibliography}
\end{document}